# A 125µW 8kS/s Sub-pA Area-Efficient Current Sensing 45nm CMOS ADC for Biosensing

Mitra Saeidi, Luke Theogarajan
University of California Santa Barbara, Santa Barbara, California, USA, mitra@ucsb.edu, lusthe@ucsb.edu

## Abstract
This paper presents a 125µW, area efficient (0.042mm$^2$) 81dB DR, 8kS/s current sensing ADC in 45nm CMOS capable of sensing sub-pA currents. Our approach combines the transimpedance amplifier (TIA) and ADC into a unified structure by folding a low-noise capacitive TIA into the first stage integrator of a 2$^{nd}$ order Delta-Sigma (ΔΣ) modulator. The dominant DAC feedback noise is mitigated by utilizing current scaling via slope modification by an integrator and differentiator pair.

## Introduction

Single-molecule electronic biosensors require high-precision current sensing (~1pA) with high bandwidth (>1kHz) and dynamic range (>10b). Higher data fidelity requires statistical averaging over a large number of sensors (10-100), placing severe constraints on the area per channel. Typical approaches require a TIA for each channel, with either capacitive or resistive feedback, followed by an ADC. Large TIA conversion gains (> 140dbΩ) required for low current sensing limits the dynamic range of this approach. An alternative approach eliminates the need for the TIA and performs direct digital conversion using current-to-digital converters thereby overcoming this limitation. Direct current-to-frequency (I-F) conversion can also improve the dynamic range but at the expense of bandwidth [1]. Delta Sigma (ΔΣ) modulators offer an alternative approach where direct current-to-digital conversion is possible [2,3]. Unfortunately, the noise of the feedback DAC fundamentally limits performance. Recently by using a 12-b SAR quantizer and a 11-b DAC a 140dB DR was achieved, since the noise of the DAC scales with the input signal [2]. However, this method comes at the expense of giving up the inherent linearity achievable in a 1-b quantizer and DAC increasing the area, complexity, and power of the converter. Our approach maintains the simplicity and linearity of the 1-b ΔΣ modulator while achieving 125dB cross-scale DR by using an advanced version of our current scaling method [3]. One major advantage of our ΔΣ approach is the ability to fold the TIA into the ADC via the first stage integrator. Another is its insensitivity to process variation due to the feedback nature of the structure. Furthermore, it allows the use of lower gain amplifiers reducing area and power. We made the following modifications to a conventional 2$^{nd}$ order ΔΣ modulator allowing us to directly sense sub-pA currents while using lower power and area. 1) A single stage common-source amplifier with a cascode load. 2) Feedback current scaling using capacitors (slope scaling, see fig. 1) enabling low feedback DAC noise and thereby enabling sub-pA sensing. 3) Correlated double sampling (CDS) to lower amplifier noise and improve process variation tolerance while increasing amplifier gain [4]. Using these techniques, we achieved a sub-pA current sensing ADC. Previously, an incremental sigma-delta using a similar capacitive current-scaling technique was proposed [5]. However, the intrinsic resetting present in the incremental approach reduces the performance.

## Architecture

Fig. 1 shows the block diagram of the wide bandwidth high precision current sensing ADC frontend. The figure also shows the slope scaling technique implemented in the feedback path, which converts the output of the current DAC into a voltage ramp via an integrator and then back into a current using a differentiator. This technique allows the scaling of the DAC current by the ratio of the integrator and differentiator capacitors. More importantly it eliminates the need for large-valued resistors normally required to lower the feedback current noise. Additionally, unlike switched capacitor feedback DACs it allows the use of small capacitors since the size does not set the integrator gain.

## Circuit Implementation

Fig. 2a shows the circuit-level block diagram of the structure. Though reminiscent of a switched capacitor ΔΣ, our approach is a continuous time (CT) ΔΣ, all the capacitor switching is opportunistic and happens during the brief reset period required for the feedback integrator. A novel correlated double sampling single stage integrator is used as main integrators in the DSM structure. The simple single stage implementation allows for lower area and power. A conventional CDS technique results in a resetting integrator, so our topology floats the integration capacitor during reset to preserve state [4]. An auxiliary capacitor charged to vcm-nbias is used to ensure the virtual ground is set to vcm, thus achieving conditions similar to a differential amplifier. The CDS technique also samples slow varying (1/f) noise and amplifier offset, improving the gain and lowering the noise. The feedback integrator required for slope-scaling reduces to a charge pump while the associated differentiator is simply a capacitor since one end is held at virtual ground by the input integrator. A differential charge pump allows for robust performance while providing the feedback signal with opposite signs making it convenient to feedback to both the first and second stage of the 2$^{nd}$ order ΔΣ. Additionally, resetting the charge pump-based integrator is fast due to its feedforward nature and only depends on the switch resistance. The fast reset times allows our design to use a relatively small fraction (1%) of the clock cycle for reset mitigating any deleterious effects posed by long reset times. Current scaling is achieved by the ratio ($\alpha$) of capacitors $C_2$ and $C_3$, i.e., $\alpha = C_3/C_2$. The current scaling also scales the noise power, $P_N = \frac{1}{\alpha^2}\overline{i_n}^2$ where $\overline{i_n}^2$ is the noise of the 1-b current DAC. Since the reference current in the DAC is scaled up by a factor of $\alpha$ to maintain the right full-scale range and noise power scales linearly with the bias current, the net noise power is reduced by $\alpha$ compared to a conventional approach. This $\alpha$ scaling allows a larger bandwidth for the same noise level. The resetting nature of the design eliminates the need for using resistors and allows for the use of a completely capacitive CT-ΔΣ design. The current design was implemented in a 45nm RF SOI process. As shown in fig. 3, the area occupied by 1 channel is (261µm×161µm), which is the smallest current sensing front-end reported (see Table I). As previously mentioned, the single stage amplifiers are common-source amplifiers with cascode loads, as shown in fig. 2b. Our current design is an amplifier with 40 dB gain effectively working as an amplifier with 80dB gain due to CDS enhancement. A variable reference current source (100nA-100µA) is designed to serve as the 1-b current DAC in the ΔΣ feedback. In the current design we chose α=100, with $C_2$=100fF and $C_3$=10pF, giving us a net reduction in current noise by a factor of 10, enabling sub-pA sensing at higher bandwidths.

## Measurement Results

Combining all these enhancements yielded a current sensing ADC with an 80dB SNDR in a 4kHz signal bandwidth with an oversampling frequency 1.024MHz and an extended signal bandwidth of 15 kHz with 72dB SNDR, as shown in fig. 4a. The SNDR performance for the complete signal range is shown in fig. 4b. The performance of the current sensing ADC translates to a FoM of 153dB. The ADC achieves this performance while consuming only 125µW/channel, with a 1V supply. The breakdown of the power is shown in fig. 5a, with the comparator consuming the most power. The sub-pA sensing capability is shown in fig. 5b, the result is for a 600fA signal at 2kHz. As shown in fig. 5c&d and Table I, our design achieves superior performance when area, power and signal bandwidth are critical. All signal measurements were performed using the nanopore circuit model shown in fig. 1.

**Acknowledgements** Special thanks to Professor Buckwalter for providing the silicon area.

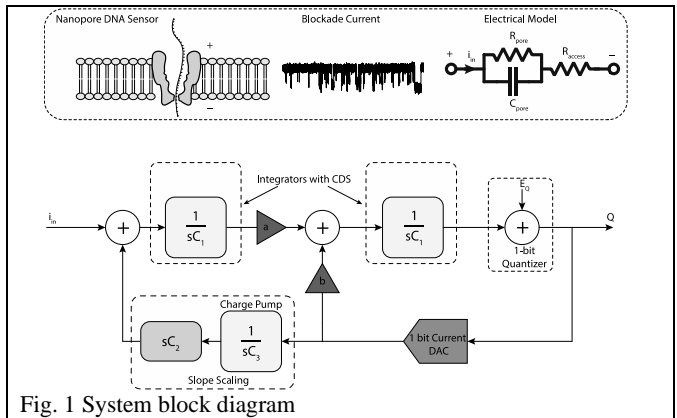

Fig. 1 System block diagram

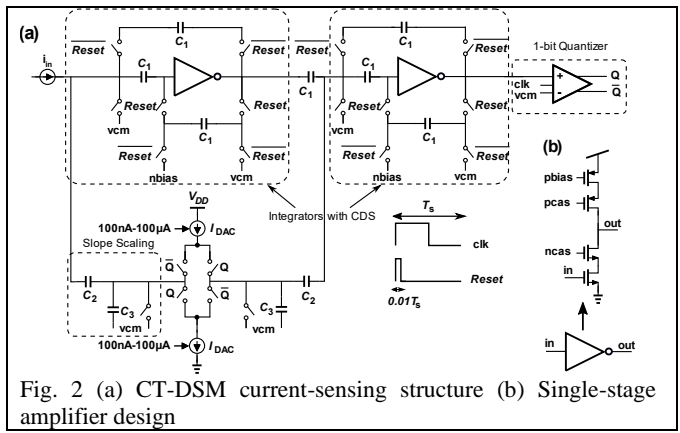

Fig. 2 (a) CT-DSM current-sensing structure (b) Single-stage amplifier design

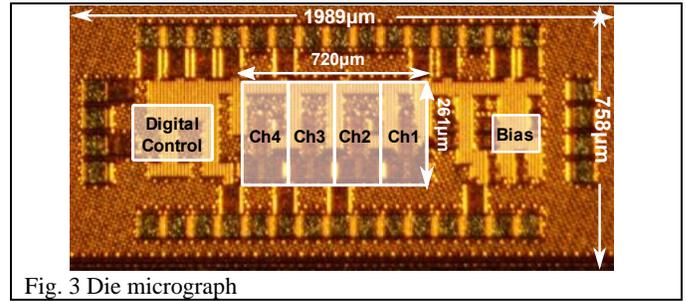

Fig. 3 Die micrograph

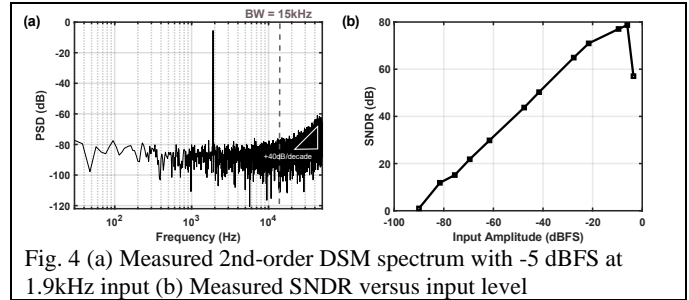

Fig. 4 (a) Measured 2nd-order DSM spectrum with -5 dBFS at 1.9kHz input (b) Measured SNDR versus input level

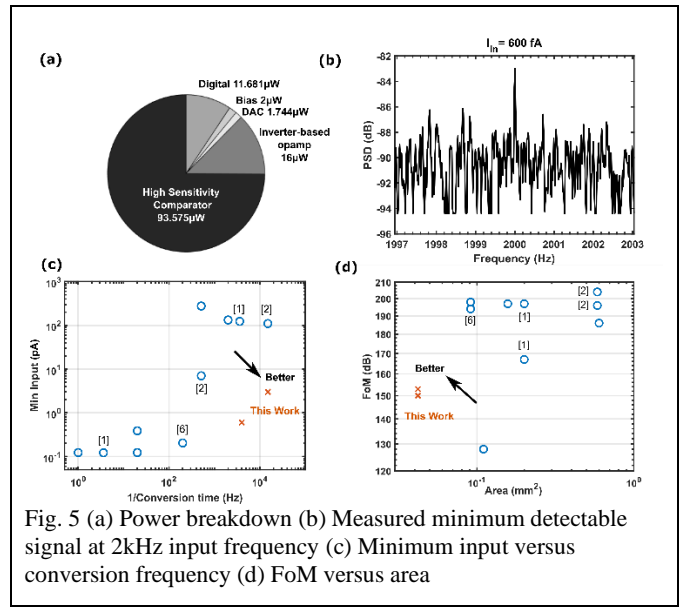

Fig. 5 (a) Power breakdown (b) Measured minimum detectable signal at 2kHz input frequency (c) Minimum input versus conversion frequency (d) FoM versus area

TABLE I

PERFORMANCE SUMMARY AND COMPARISON

|  | This Work | [1] | [2] | [6] |
|---|---|---|---|---|
| Process [nm] | 45 | 180 | 55 | 180 |
| Power/channel [µW] | 125 | 295 | 1011 | 5220 |
| Supply [v] | 1 | 1.8 | 1.2 | 1.8 |
| Number of Channels | 4 | 1 | 1 | 1 |
| Active Area/channel [mm²] | 0.042 | 0.2 | 0.585 | 0.091 |
| Max input measured | 1µA | 10µA | 200µA | 11.6µA |
| Reference current for min input | 10nA | 10µA | 200µA | 11.6µA |
| Min Input @ $F_{conv}$ [pA] | 0.6 @4kHz | 123@ 3.6kHz 0.122@3.6kHz | 25 @ 4kHz 7.75@512Hz | 0.204@200Hz |
| Fixed/ Cross-scale Dynamic Range @ BW [dB] | 72/112.2@15kHz 81/125@4kHz | 100@3.6KHz 120@360Hz 140@36Hz 160@3.6Hz | 127@15KHz 136@7.8KHz 140@4KHz 150@512Hz | 155@200Hz |
| Fixed/ Cross-scale $FoM_{Schreier}$[dB] | 150/190@15kHz 153/197@4kHz | 167@3.6KHz 177@360Hz 187@36Hz 197@3.6Hz | 196@15KHz 202@7.8KHz 203@4KHz 204@512Hz | 194@200Hz |

(Fixed/ Cross-scale) $FoM_{Schreier}$ [dB] = $DR_{(Fixed/\ Cross-scale)}$ [dB]+10 $log_{10}$($F_{conv}$/2/Power)